\documentclass[10pt,conference]{IEEEtran}

\usepackage{cite}
\usepackage{graphicx}
\usepackage{psfrag}
\usepackage{subfigure}
\usepackage{flushend}
\usepackage{url}
\usepackage[cmex10]{amsmath}
\usepackage{array}
\usepackage{graphicx}
\usepackage{epsfig}
\usepackage{amsbsy}
\usepackage{amssymb}
\usepackage{multicol}
\usepackage{amsthm}
\usepackage{times}

\newcounter{MYtempeqncnt}
\IEEEoverridecommandlockouts

\begin{document}

\title{\LARGE On the Effect of Primary User Traffic on Secondary Throughput and
Outage Probability Under Rayleigh Flat Fading Channel}
\author{\IEEEauthorblockN{Sanket S. Kalamkar and Adrish Banerjee\\
Department of Electrical Engineering, Indian Institute of Technology Kanpur, India
\thanks{This work was supported in part by a research grant from the Indo-
UK Advanced Technology Centre and a research scholarship from the Tata
Consultancy Services Research Scholar Program.}}
Email: \{kalamkar, adrish\}@iitk.ac.in}
\maketitle

\begin{abstract}
The effect of the primary traffic on a secondary user's (SU) throughput under Rayleigh flat fading channel is investigated. For this case, closed form expressions are derived for the average probability of detection and the average probability of false alarm. Based on these expressions, the average SU throughput under the desired signal-to-noise ratio (SNR) constraint in order to maintain the quality of the secondary link is found analytically considering the random arrival or departure of the primary user. It is shown that the spectrum sensing performance and SU throughput degrade with increase in the primary traffic and the deep fade condition of the channel over which the detection is performed. The degree of degradation in SU throughput is seen to be severed further due to the interference link from the primary transmitter to the secondary receiver. Under these detrimental effects, a sensing-throughput trade-off for SU is illustrated. Finally, the combined effect of the primary traffic, fading, imperfect spectrum sensing and the interference link from a primary transmitter is studied on the outage probability at SU.   
\end{abstract}
\begin{IEEEkeywords}
Cognitive radio, interference, outage probability, primary user traffic, Rayleigh fading channel, spectrum sensing.
\end{IEEEkeywords} \vspace*{-2mm}

\section{Introduction}
Spectrum sensing plays an important role in cognitive radio. An unlicensed or a secondary user (SU) performs spectrum sensing to determine the occupancy of a licensed band that belongs to a licensed user or a primary user (PU). Different spectrum sensing techniques like energy detection, cyclostationary detection and matched filter detection \cite{yucek} have been proposed in literature. Energy detection \cite{urkowitz} is the most commonly used spectrum sensing technique due to its low complexity and it does not need any knowledge about the primary signal. Periodic sensing is necessary to know the occupancy of the channel on a regular basis. A longer sensing period leads to a better spectrum sensing performance. However, increase in the sensing period reduces the transmission period available for secondary transmission for a fixed frame duration, in turn, reducing the throughput of SU\cite{liang}.  

In \cite{liang}, the sensing-throughput trade-off is studied for a target probability of detection $P_D$. However, it is assumed that PU remains constantly  either present or absent in the channel during the entire frame duration of SU. In \cite{solta, shar, juarez, juarez1}, the sensing-throughput trade-off is studied for fading channels. But, all these works did not take into consideration the primary traffic. In practice, PU may randomly arrive or leave the licensed band at any time during the frame duration. This is the case especially when PU network has a high traffic rate or when the frame duration of SU is long. The random nature of the primary traffic affects the throughput of SU. This effect is intense especially when PU is absent in the sensing period and arrives during the transmission of SU causing interference to SU. Hence, the study of sensing-throughput trade-off under primary traffic is of great interest. This is addressed in \cite{ychen1, lu} for an additive white Gaussian noise (AWGN) channel, but the fading effect is not taken into account. The random nature of the fading channel may add up to the random arrival and departure of PU, which not only makes the spectrum sensing more difficult but also significantly reduces the achievable throughput of the secondary link especially, if the channel is in deep fade. In \cite{chen2}, the fading channel is considered under the primary traffic. However, the sensing-throughput trade-off and the outage analysis are missing from \cite{chen2}. 

The contributions of this paper are as follows:
\begin{itemize}
\item Firstly, we consider a scenario where SU performs detection and transmission over Rayleigh flat fading channels under random primary traffic. For this scenario, we derive closed form expressions for the average probability of detection and the average probability of false alarm.
\item Secondly, we put the desired SNR constraint on SU's transmission to ensure the reliable communication. Under this constraint, we analytically characterize the effect of Rayleigh fading and the primary traffic on the average throughput of SU.
\item  Thirdly, using numerical simulations, the sensing-throughput trade-off of SU is studied considering an interference link from PU transmitter to SU receiver due to imperfect sensing and the arrival of PU in the transmission duration of SU. 
\item Fourthly, an expression for the outage probability at SU for the desired SNR constraint is obtained analytically. Furthermore, the combined effect of the primary traffic, fading, primary interference and imperfect sensing on the outage probability is illustrated.
\end{itemize}
 
\section{System Model}
\subsection{Channel model}
Consider a secondary link consisting of a SU transmitter ($ST$) and a SU receiver ($SR$). There exists a PU transmitter ($PT$) to which the licensed channel of SU's interest belongs\footnote[1]{Since the focus of this paper to study SU throughput against the primary traffic for fading channels, we do not include a PU receiver in the system model.}. All channels are considered Rayleigh flat fading channels. Let us denote the instantaneous channel power gains between $PT \rightarrow ST$, $PT \rightarrow SR$ and $ST \rightarrow SR$ by  $|h|^2 \sim \exp(\lambda_h)$, $|\chi|^2 \sim \exp(\lambda_{\chi})$ and $|g|^2 \sim \exp(\lambda_g)$, respectively, where $\exp(k)$ is an exponential distribution with parameter $k$ and mean value $\frac{1}{k}$. It is assumed that the average channel power gains are available to SU. 
\subsection{Spectrum sensing under no primary traffic (Conventional model)}
In conventional model (with no primary traffic), it is assumed that PU is either present or absent during the entire frame duration. Then, the energy detection of PU can be modelled as a binary hypothesis problem as 
\begin{equation}
\label{eq:binary}
 Y_{con} = \left\{
  \begin{array}{l l}
    \sum_{n = 1}^{N}(hs_n + w_n)^{2}, & \quad H_1, \\
    \sum_{n = 1}^{N}{w_n}^{2}, & \quad H_0,\\
  \end{array} \right.
\end{equation}
where $Y_{con}$ is the output of an energy detector, $N$ is the total number of sensing samples, $s_n$ is $n$th sample of PU signal, $w_n$ is i.i.d. mean zero AWGN with variance $\sigma_{ST}^2$, $H_1$ and $H_0$ are the hypotheses corresponding to the presence and absence of PU, respectively. Then, the average probability of false alarm ($P_{F}$) and detection ($P_{D}$) are given as \cite{digham}
\begin{eqnarray}
\label{eq:pf_con}
P_{F} = \frac{\Gamma(u, \eta/2)}{\Gamma(u)},
\end{eqnarray}\vspace*{-3mm}
\begin{eqnarray}
\label{eq:pd_con}
P_{D} &=& \int_0^{\infty} \underbrace{\hbox{$Q_u(\sqrt{2u\gamma}, \sqrt{\eta})$}}_{\hbox{$P_{D,AWGN}$}}f(\gamma)\,\mathrm{d}\gamma = e^{-\frac{\eta}{2}}\sum_{i = 0}^{u - 2}\frac{1}{i!}\left( \frac{\eta}{2}\right)^{i} \nonumber \\ 
&&\hspace*{-18mm}+ \left(\frac{1 + u\bar{\gamma}_p}{u\bar{\gamma}_p}\right)^{u-1}\left[e^{-\frac{\eta}{2(1 + u\bar{\gamma}_p)}} - e^{-\frac{\eta}{2}}\sum_{i = 0}^{u - 2}\frac{1}{i!}\left(\frac{\eta u\bar{\gamma}_p}{2(1 + u\bar{\gamma}_p)}\right)^{i}\right],\nonumber \\
\end{eqnarray}
where $\eta$ is the detection threshold, $u = N/2$ is the time-bandwidth product, $\Gamma(\cdot,\cdot)$ and $\Gamma(\cdot)$ are the incomplete and complete gamma functions\cite{gradshteyn}, $Q_u(\cdot, \cdot)$ is the generalized Marcum $Q$-function \cite{nuttall}, $P_{D,AWGN}$ is the probability of detection for AWGN channel and $f(\gamma)$ is the probability density function of SNR $\gamma$. When the channel is Rayleigh, $f(\gamma)$ is given as 
\begin{equation}
f(\gamma) = \frac{1}{\bar{\gamma}_p}\exp\left(-\frac{\gamma}{\bar{\gamma}_p}\right), \hspace*{4mm} \gamma \geq 0,
\label{eq:snr_ray}\vspace*{-1mm}
\end{equation}
with $\bar{\gamma}_p = \frac{P_p}{\lambda_h\sigma_{ST}^2}$ as the average received primary SNR at $ST$ where $P_p$ is PU's transmit power. Thus, the probability of detection for Rayleigh fading channel is obtained by averaging the probability of detection for AWGN channel over \eqref{eq:snr_ray}. 

\subsection{Spectrum sensing under primary traffic}
We model the behavior of PU as a busy-idle random process, where busy ($B$) and idle ($I$) represent the presence and absence of PU, respectively. The busy and idle periods are assumed to be exponentially distributed random variables with mean parameters $\alpha$ for a busy state and $\beta$ for an idle state. Then, the stationary probabilities for the busy state of PU is given by $P_B = \frac{\alpha}{\alpha + \beta}$ and for the idle state by $P_I = \frac{\beta}{\alpha + \beta}$. The traffic parameters $\alpha$ and $\beta$ can be estimated using statistical methods \cite{fzhang}. Given PU was in state `$\omega \in \lbrace B, I\rbrace$' $t$ seconds ago, PU now appears in state `$\Omega \in \lbrace B, I\rbrace$' with the transition probability matrix \cite{papoulis} 
\begin{eqnarray}
P_{\omega \Omega}(t) \hspace*{-2mm}&=&\hspace*{-2mm}
 \begin{pmatrix}
  P_{II} & P_{IB} \\
  P_{BI} & P_{BB}  
 \end{pmatrix} \nonumber \\
\hspace*{-2mm} &=& \hspace*{-2mm}\frac{1}{\alpha + \beta} \begin{pmatrix}
  \alpha + \beta e^{-(\alpha + \beta)t} & \beta - \beta e^{-(\alpha + \beta)t} \\
  \alpha - \alpha e^{-(\alpha + \beta)t} & \beta + \alpha e^{-(\alpha + \beta)t} 
 \end{pmatrix}.
 \end{eqnarray}
We consider a scenario where PU changes its status at most once as the probability that PU changes status at most once in a frame is large \cite{ychen1, lu}. This assumption is reasonable for the practical cases where $MT_\text{samp}$ $\ll$ \text{min}$\left[\frac{1}{\alpha}, \frac{1}{\beta}\right]$, where $M$ is the total number of samples in a frame of fixed duration and $T_\text{samp}$ is the sample interval. We consider the case where PU can change its status any time in a frame and takes one sample for the transition. Let $a$ and $d$ denote the sample after which PU arrives or departs from the licensed channel, respectively. 

Under primary traffic and based on the above model, PU detection problem is now an quaternary hypothesis testing problem and is given as \cite{ychen1} 
\begin{equation}
\label{eq:quaternary}
 Y \hspace*{-1mm}=\hspace*{-1mm} \left\{
  \begin{array}{l l}
    \sum_{n = 1}^{N}(hs_n + w_n)^{2}, & \quad H_{1,1}, \\
    \sum_{n = 1}^{a}{w_n}^{2} + \sum_{n = a+1}^{N}(hs_n + w_n)^{2}, & \quad H_{1,2},\\
    \sum_{n = 1}^{N}{w_n}^{2}, & \quad H_{0,1},\\
     \sum_{n = 1}^{d}(hs_n + w_n)^{2} + \sum_{n = d + 1}^{N}{w_n}^{2}, & \quad H_{0,2},
  \end{array} \right.
\end{equation}
where the hypothesis $H_{1,1}$ corresponds to the presence of PU during the entire sensing period (same as hypothesis $H_1$ in the model with no primary traffic given by \eqref{eq:binary}); the hypothesis $H_{1,2}$ is that PU is absent at the beginning of the sensing period for $a$ samples and then arrives in the licensed band; the hypothesis $H_{0,1}$ corresponds to the absence of PU during the entire sensing period (same as hypothesis $H_0$ in the model with no primary traffic); the hypothesis $H_{0,2}$ is that PU is present at the beginning of the sensing period for $d$ samples and then departs from the licensed band. Then, the probability of each hypothesis can be written as \cite{ychen1}
\begin{eqnarray}
\label{eq:prob_h}
P(H_{1,1})\hspace*{-1mm} &=& \hspace*{-2mm}P_BP_{BB}^{M} + \sum_{d = N + 1}^{M - 1}\left[P_B(P_{BB})^{d}P_{BI}(P_{II})^{M - d -1}\right], \nonumber \\
P(H_{1,2})&=&\sum_{a = 1}^{N}\left[P_I(P_{II})^{a}P_{IB}(P_{BB})^{M - a -1}\right],\nonumber\\
P(H_{0,1}) &=&P_IP_{II}^{M} + \sum_{a = N + 1}^{M - 1}\left[P_I(P_{II})^{a}P_{IB}(P_{BB})^{M - a -1}\right],\nonumber \\
P(H_{0,2}) &= &\sum_{d = 1}^{N}\left[P_B(P_{BB})^{d}P_{BI}(P_{II})^{M - d -1}\right].
\end{eqnarray}
Based on \eqref{eq:quaternary}, the conditional probability of detection and the conditional probability of false alarm can be obtained for each hypothesis. For $H_{1,1}$, PU remains present during the entire sensing period. Thus, the conditional probability of detection is same as \eqref{eq:pd_con} and is given by 

{{\small{
\begin{eqnarray}
\label{eq:pd1_new}
P_{DH_{1,1}}(u) &=& Pr(H_1|H_{1,1}) = Pr(Y > \eta|H_{1,1}) \nonumber \\
&\hspace*{-35mm}=&\hspace*{-20mm}\int_0^{\infty} \underbrace{\hbox{$Q_u(\sqrt{2u\gamma}, \sqrt{\eta})$}}_{\hbox{$P_{D,AWGN}$}}f(\gamma)\,\mathrm{d}\gamma  = e^{-\frac{\eta}{2}}\sum_{i = 0}^{u - 2}\frac{1}{i!}\left( \frac{\eta}{2}\right)^{i} \nonumber \\
&&\hspace*{-27mm}+ \left(\frac{1 + \bar{\mu}_p}{\bar{\mu}_p}\right)^{u-1} \left[e^{-\frac{\eta}{2(1 + \bar{\mu}_p)}} - e^{-\frac{\eta}{2}}\sum_{i = 0}^{u - 2}\frac{1}{i!}\left(\frac{\eta \bar{\mu}_p}{2(1 + \bar{\mu}_p)}\right)^{i}\right],\vspace*{-1mm}
\end{eqnarray}}}}\vspace*{-1mm}

\noindent where $\bar{\mu}_p = u\bar{\gamma}_p$. For $H_{1,2}$, PU arrives in the sensing period after $a$ samples ($0 < a \leq N$). Thus, SU has to detect PU arrival by collecting $2u - a$ samples over the remaining sensing period. Thus, the conditional probability of detection conditioned on $a$ is given by making use of \cite[Eq. (30)]{nuttall} as \vspace*{-3mm}
 
 {{\small{
\begin{eqnarray}
\label{eq:pd2_new}
\hspace*{-4mm}P_{DH_{1,2}}(u, a) &=& Pr(H_1|H_{1,2}, a) = Pr(Y > \eta|H_{1,2}, a) \nonumber \\
&\hspace*{-43mm}=&\hspace*{-23mm} \int_0^{\infty} \underbrace{\hbox{$Q_u(\sqrt{(2u - a)\gamma}, \sqrt{\eta})$}}_{\hbox{$P_{D,AWGN}$}}f(\gamma)\,\mathrm{d}\gamma = e^{-\frac{\eta}{2}}\sum_{i = 0}^{u - 2}\frac{1}{i!}\left( \frac{\eta}{2}\right)^{i} \nonumber \\
&&\hspace*{-27mm}+ \left(\frac{1 + \bar{\delta}_p}{\bar{\delta}_p}\right)^{u-1}\left[e^{-\frac{\eta}{2(1 + \bar{\delta}_p)}} - e^{-\frac{\eta}{2}}\sum_{i = 0}^{u - 2}\frac{1}{i!}\left(\frac{\eta \bar{\delta}_p}{2(1 + \bar{\delta}_p)}\right)^{i}\right],
\end{eqnarray}}}}\vspace*{-1mm}

\noindent where $\bar{\delta}_p = \frac{(2u - a)}{2}\bar{\gamma}_p$.
For $H_{0,1}$, PU remains absent during the entire sensing duration. Thus, the conditional probability of false alarm is same as \eqref{eq:pf_con} and is given by\vspace*{-2mm}
\begin{eqnarray}
\label{eq:pf1_new}
P_{FH_{0,1}}(u) &=& Pr(H_1|H_{0,1}) = Pr(Y > \eta|H_{0,1})  \nonumber\\
&=&\frac{\Gamma(u, \eta/2)}{\Gamma(u)}.
\end{eqnarray}
For $H_{0,2}$, PU departs in the sensing period after $d$ samples ($0 < d \leq N$). Thus, the conditional probability of false alarm conditioned on $d$ can be given by making use of \cite[Eq. (30)]{nuttall} as \vspace*{-2mm}

{{\small{
\begin{eqnarray}
\label{eq:pf2_new}
\hspace*{-4mm}P_{FH_{0,2}}(u, d)&=& Pr(H_1|H_{0,2}, d) = Pr(Y > \eta|H_{0,2}, d) \nonumber \\
&\hspace*{-43mm}=&\hspace*{-23mm} \int_0^{\infty} \underbrace{\hbox{$Q_u(\sqrt{d\gamma}, \sqrt{\eta})$}}_{\hbox{$P_{D,AWGN}$}}f(\gamma)\,\mathrm{d}\gamma = e^{-\frac{\eta}{2}}\sum_{i = 0}^{u - 2}\frac{1}{i!}\left( \frac{\eta}{2}\right)^{i} \nonumber \\
&&\hspace*{-27mm}+ \left(\frac{1 + \bar{\rho}_p}{\bar{\rho}_p}\right)^{u-1}\left[e^{-\frac{\eta}{2(1 + \bar{\rho}_p)}} - e^{-\frac{\eta}{2}}\sum_{i = 0}^{u - 2}\frac{1}{i!}\left(\frac{\eta \bar{\rho}_p}{2(1 + \bar{\rho}_p)}\right)^{i}\right],
\end{eqnarray}}}}\vspace*{-1mm}

\noindent where $\bar{\rho}_p = \frac{d}{2}\bar{\gamma}_p$. It can be seen that putting $a = 0$ in \eqref{eq:pd2_new} and $d = 0$ in \eqref{eq:pf2_new}, we get \eqref{eq:pd_con} and \eqref{eq:pf_con}, respectively.

Then, the unconditional probability of detection and false alarm can be obtained by averaging out conditional probabilities over the probabilities of hypotheses and are given by
\begin{eqnarray}
\label{eq:pd_avg}
\overline{P_{D}} &=& \frac{P(H_{1,1})P_{DH_{1,1}}(u)}{P(H_{1,1}) + P(H_{1,2})} \nonumber\\
&& \hspace*{-15mm}+ \sum_{a = 1}^{N} \left(\frac{P_I(P_{II})^{a}P_{IB}(P_{BB})^{M - a -1}}{P(H_{1,1}) + P(H_{1,2})} P_{DH_{1,2}}(u, a)\right),
\end{eqnarray}\vspace*{-1mm}
\begin{eqnarray}
\label{eq:pf_avg}
\overline{P_{F}} &=& \frac{P(H_{0,1})P_{FH_{0,1}}(u)}{P(H_{0,1}) + P(H_{0,2})} \nonumber\\
&& \hspace*{-15mm}+ \sum_{d = 1}^{N} \left(\frac{P_B(P_{BB})^{d}P_{BI}(P_{II})^{M - d -1}}{P(H_{0,1}) + P(H_{0,2})} P_{FH_{0,2}}(u, d)\right).
\end{eqnarray}
\section{Secondary throughput analysis under primary traffic and flat fading Rayleigh channel}\vspace*{-1mm}

In this section, we analytically find the average throughput of SU over the secondary link ($ST \rightarrow SR$) for flat fading Rayleigh channel with the primary traffic under the desired SNR constraint. Also, an expression for the outage probability at SU is found. \vspace*{-1mm}
\subsection{Secondary Throughput Analysis}
The secondary user may start its transmission once it detects the licensed band idle after performing spectrum sensing as in \eqref{eq:quaternary}. When PU is absent, the probability of successful transmission over the secondary link while maintaining the desired SNR above a threshold $\gamma_s$ can be given by
\begin{equation}
\label{eq:th_puas}
Pr\left(\frac{P_s|g|^2}{\sigma_{SR}^2} > \gamma_s\right) = \exp\left(-\frac{\sigma_{SR}^2\lambda_g \gamma_s}{P_s}\right),
\end{equation}
where $\gamma_s$ denotes SU's SNR threshold (or SINR, i.e., signal-to-interference noise ratio if PU is present), $\sigma_{SR}^2$ is noise variance at SU receiver.  and $P_s$ is SU's transmit power. Here, since SU transmission is to be performed with the minimum SNR $\gamma_s$, $\text{log}_2(1 + \gamma_s)$ becomes the minimum transmission rate obtained in the absence of the primary traffic, fading, primary interference and imperfect sensing as given in \cite{shar}. Then, the minimum local SU throughput that can be obtained becomes\vspace*{-1mm}
\begin{equation}
\label{eq:th_pua}
L_{H_0} = \exp\left(-\frac{\sigma_{SR}^2\lambda_g \gamma_s}{P_s}\right)\text{log}_2(1 + \gamma_s).
\end{equation} 
When PU is present, the probability of successful transmission over the secondary link not only depends on the fading condition of the secondary link but also on the primary interference and can be given by\vspace*{-1mm}
\begin{eqnarray}
\label{eq:pu_p}
&&Pr\left(\frac{P_s|g|^2}{\sigma_{SR}^2 + P_p|\chi|^2} > \gamma_s\right) \nonumber \\
&=&\mathbb{E}_{|\chi|^2}\bigg(\int_{(\sigma_{SR}^2 + P_p|\chi|^2)\gamma_s/P_s}^{\infty} \lambda_g \exp\left(-\lambda_g|g|^2\right)\,\mathrm{d}|g|^2 \bigg) \nonumber \\
&=& \frac{\lambda_{\chi}\exp\left(-\frac{\sigma_{SR}^2\lambda_g \gamma_s}{P_s}\right)}{\lambda_{\chi} + \frac{\lambda_g\gamma_sP_p}{P_s}}.
\end{eqnarray}
The equation \eqref{eq:pu_p} is obtained by averaging out the probability density function of the channel power gain $|g|^2$ over all the possible channel realizations of the interference link from the primary transmitter. Then, the minimum local SU throughput in the presence of PU is given by \vspace*{-3mm}

\begin{equation}
\label{eq:th_pre}
L_{H_1} = \frac{\lambda_{\chi}\exp\left(-\frac{\sigma_{SR}^2\lambda_g \gamma_s}{P_s}\right)}{\lambda_{\chi} + \frac{\lambda_g\gamma_sP_p}{P_s}} \text{log}_2(1 + \gamma_s).
\end{equation}
Now, for $H_{1,1}$ in \eqref{eq:quaternary}, PU is always present during the entire sensing period and it may stay there or leave the licensed band during the transmission period. The minimum local SU throughput for this case can be given by
\begin{figure*}
\normalsize
\setcounter{MYtempeqncnt}{\value{equation}}
\setcounter{equation}{22}
\begin{eqnarray}
\label{eq:thr_exp11}
R_{H_{1,1}} & =& \frac{M - N}{M}\left(1 - \overline{P_{D}}\right)\bigg(P_BP_{BB}^{M}L_{H_{1,1}}(M) + \sum_{d = N + 1}^{M - 1}\big(P_B(P_{BB})^{d}P_{BI}(P_{II})^{M - d -1}L_{H_{1,1}}(d)\big)\bigg),
\end{eqnarray} \vspace*{-5mm}
\begin{eqnarray}
\label{eq:thr_exp12}
\hspace*{-39mm} R_{H_{1,2}} &=& \frac{M - N}{M}\left(1 - \overline{P_{D}}\right)
 \sum_{a = 1}^{N}\bigg(P_I(P_{II})^{a}P_{IB}(P_{BB})^{M - a-1}L_{H_{1,2}}\bigg),
\end{eqnarray}\vspace*{-5mm}
\begin{eqnarray}
\label{eq:thr_exp01}
\hspace*{-7mm}R_{H_{0,1}} & =& \frac{M - N}{M}\left(1 - \overline{P_{F}}\right)\bigg(P_IP_{II}^{M}L_{H_{0,1}}(M) 
 \sum_{a = N + 1}^{M - 1}\big(P_I(P_{II})^{a}P_{IB}(P_{BB})^{M - a -1}L_{H_{0,1}}(a)\big)\bigg),
\end{eqnarray}\vspace*{-5mm}
\begin{eqnarray}
\label{eq:thr_exp02}
\hspace*{-39mm}R_{H_{0,2}} &=& \frac{M - N}{M}\left(1 - \overline{P_{F}}\right) 
\sum_{d = 1}^{N}\bigg(P_B(P_{BB})^{d}P_{BI}(P_{II})^{M - d-1}L_{H_{0,2}}\bigg).
\end{eqnarray}
\setcounter{equation}{\value{MYtempeqncnt}}
\hrulefill
\vspace*{-5mm}
\end{figure*}
\begin{eqnarray}
\label{eq:th_H11}
L_{H_{1,1}}(d) &=& \left(\frac{d - N}{M - N}\right)L_{H_{1}} +  \left(\frac{M - d}{M - N}\right)L_{H_{0}}
\end{eqnarray}
for $N+1 \leq d \leq M$. $L_{H_1}$ and $L_{H_0}$ are given by \eqref{eq:th_pre} and \eqref{eq:th_pua}, respectively. It can be seen that from \eqref{eq:th_H11} that the probability of successful transmission depends on when PU departs from the licensed band. In $H_{1,2}$, PU arrives during the sensing period and it stays present for rest of the frame. Then, the minimum local SU throughput obtained is same as \eqref{eq:th_pre} and is given by
\begin{equation}
\label{eq:th_H12}
L_{H_{1,2}} = \frac{\lambda_{\chi}\exp\left(-\frac{\sigma_{SR}^2\lambda_g \gamma_s}{P_s}\right)}{\lambda_{\chi} + \frac{\lambda_g\gamma_sP_p}{P_s}}\text{log}_2(1 + \gamma_s).
\end{equation}
In $H_{0,1}$, PU is absent during the entire sensing period and it may stay absent or arrive during the transmission period. Then, the minimum local SU throughput becomes,\vspace*{-1mm} 
\begin{eqnarray}
\label{eq:th_H01}
L_{H_{0,1}}(a) &=&\left(\frac{M - a}{M - N}\right)L_{H_{1}} +  \left(\frac{a - N}{M - N}\right)L_{H_{0}}\vspace*{-1mm} 
\end{eqnarray}
for $N+1 \leq a \leq M$. $L_{H_1}$ and $L_{H_0}$ are given by \eqref{eq:th_pre} and \eqref{eq:th_pua}, respectively. In \eqref{eq:th_H01}, the probability of successful transmission depends on when PU arrives in the licensed band. In $H_{0,2}$, PU departs the licensed channel during sensing period and stays idle for rest of the frame. Then, the minimum local SU throughput is same as \eqref{eq:th_pua} and is given by\vspace*{-1mm} 
\begin{equation}
\label{eq:th_H02}
L_{H_{0,2}} = \exp\left(-\frac{\sigma_{SR}^2\lambda_g \gamma_s}{P_s}\right)\text{log}_2(1 + \gamma_s).
\end{equation} 
One can notice that if $a = N$ or $a = M$, \eqref{eq:th_H01} degenerates to \eqref{eq:th_H12} or \eqref{eq:th_H02}, respectively. Similarly, if $d = N$ or $d = M$, \eqref{eq:th_H11} degenerates to \eqref{eq:th_H02} or \eqref{eq:th_H12}, respectively.

Using \eqref{eq:prob_h}, \eqref{eq:pd_avg}, \eqref{eq:pf_avg}, \eqref{eq:th_H11}, \eqref{eq:th_H12}, \eqref{eq:th_H01} and \eqref{eq:th_H02}, the minimum average throughput of SU can be obtained as\vspace*{-2mm} 
\begin{equation}
\label{eq:thr}
R = R_{H_{1,1}} + R_{H_{1,2}} + R_{H_{0,1}} + R_{H_{0,2}},\vspace*{-2mm} 
\end{equation}
where $R_{H_{1,1}}$, $R_{H_{1,2}}$, $R_{H_{0,1}}$ and $R_{H_{0,2}}$ are given by \eqref{eq:thr_exp11}, \eqref{eq:thr_exp12}, \eqref{eq:thr_exp01} and \eqref{eq:thr_exp02}, respectively. 

\begin{itemize}
\item For $H_{1,1}$, there are two possibilities regarding PU's status in the transmission period of SU. Firstly, PU can stay present during the entire transmission period of SU. This possibility corresponds to the first term in the big bracket of \eqref{eq:thr_exp11}. Secondly, PU may leave after $d$ samples of the transmission period ($N+1 \leq d \leq M-1$), which corresponds to the summation term of \eqref{eq:thr_exp11}. Here, one can note that $N+1 \leq d \leq M-1$ is the case when PU leaves during the transmission period of SU, while $d = M$ indicates the case when PU remains present during the entire transmission period of SU. Thus, $R_{H_{1,1}}$ is calculated considering the local throughput in each event times the probability of the corresponding event summing over all the possible departure instances of PU in the transmission period. Also, $\frac{M-N}{M}$ denotes the fraction of the total frame duration used as the transmission period of SU, while $1 - \overline{P_D}$ is the probability with which SU fails to detect the presence of PU in the sensing period and starts its transmission over the licensed band.\vspace*{1mm}
\item In $H_{1,2}$, since PU arrives in the sensing period after $a$ samples ($1 \leq a \leq N$), it stays present during the entire transmission period of SU (since PU changes its status at most once during a frame duration of SU). In this case, $R_{H_{1,2}}$ given by \eqref{eq:thr_exp12} becomes the local SU throughput for $H_{1,2}$ times the probability of the hypothesis $H_{1,2}$ summing over all the possible arrival instances of PU in the sensing period.\vspace*{1mm}
\item For $H_{0,1}$, there exists two possibilities regarding PU's status in the transmission period of SU. Firstly, PU can stay absent during the entire transmission period of SU. This possibility corresponds to the first term in the big bracket of \eqref{eq:thr_exp01}. Secondly, PU may arrive after $a$ samples of the transmission period ($N+1 \leq a \leq M-1$), which corresponds to the summation term of \eqref{eq:thr_exp01}. Here, one can note that $N+1 \leq a \leq M-1$ is the case when PU arrives during the transmission period of SU, while $a = M$ indicates the case when PU remains absent during the entire transmission period of SU. Thus, $R_{H_{0,1}}$ is calculated considering the local throughput in each event times the probability of the corresponding event summing over all the possible arrival instances of PU in the transmission period. Also, $1 - \overline{P_F}$ is the probability with which SU succeeds to detect the absence of PU in the sensing period and starts its transmission over the licensed band.\vspace*{1mm}
\item In $H_{0,2}$, since PU departs in the sensing period after $d$ samples ($1 \leq d \leq N$), it stays absent during the entire transmission period of SU (since PU changes its status at most once during a frame duration of SU). In this case, $R_{H_{0,2}}$ given by \eqref{eq:thr_exp02} becomes the local SU throughput for $H_{0,2}$ times the probability of the hypothesis $H_{0,2}$ summing over all the possible departure instances of PU in the sensing period.
\end{itemize}

\subsection{Outage Probability at SU}

From \eqref{eq:thr}, it can be seen that $R$ is the minimum achievable rate, which is obtained by multiplying the probability of successful transmission $P_{succ}$ considering all the four hypothesis by the minimum transmission rate $\log_2({1 +\gamma_s})$ (as $\gamma_s$ is the minimum SNR required) achieved when there is no failed transmission in the absence of fading, primary interference and primary traffic. Thus, the probability of successful transmission considering all the four hypotheses under mentioned detrimental effects is given by $\frac{R}{\log_2(1 + \gamma_s)}$. Then, the outage probability becomes 
\begin{eqnarray}
\text{Outage Probability} &=& 1-P_{succ} \nonumber \\
&=& 1 - \frac{R}{\log_2(1 + \gamma_s)}.
\end{eqnarray}

\section{Numerical Results and Discussions}

In this section, we discuss numerical results showing the effect of various factors like the primary traffic, fading, primary interference, imperfect sensing and the SNR constraint on SU throughput and the outage probability at SU. The following parameters are considered without loss of generality to obtain numerical results: Sampling time, ${T_{\text{samp}}} = \mathrm{0.1ms}$, sensing duration, $T_s = \mathrm{5ms}$, frame duration, $T_{F} = 25 \mathrm{ms}$, $\lambda_g = 1$, $P_s$ = $\mathrm{1}$, $\gamma_s = 5\mathrm{dB}$, target probability of detection, $\overline{P_D} = 0.9$. Also, $\mathrm{10}\log_{10}\frac{P_p}{\lambda_h \sigma_{ST}^2}\bigg|_{P_p = 1}$ = 5dB and $\mathrm{10}\log_{10}\frac{P_s}{\lambda_g \sigma_{SR}^2}$ = 20dB. The sensing period is denoted by $T_s$.

 \begin{figure}
 \centering 
\includegraphics[scale=0.4]{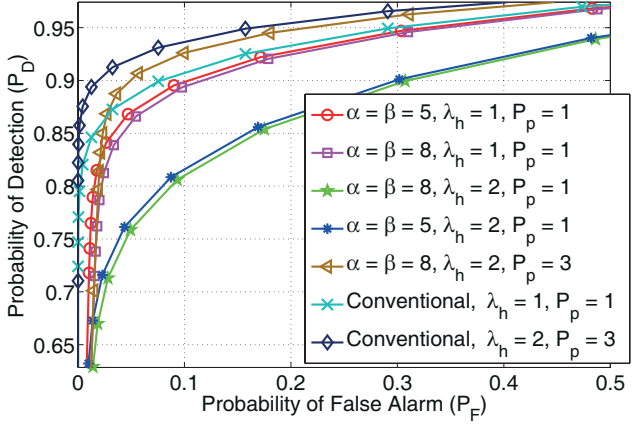} \vspace*{-7mm}
 \caption{Effect of primary traffic and fading on the detection performance of SU, $\lambda_{\chi}$ = $\mathrm{1}$, (zoomed-in view).}
      \label{fig:pd_pf}\vspace*{-4mm}
          \end{figure}
          
    \begin{figure}
 \centering 
\includegraphics[scale=0.4]{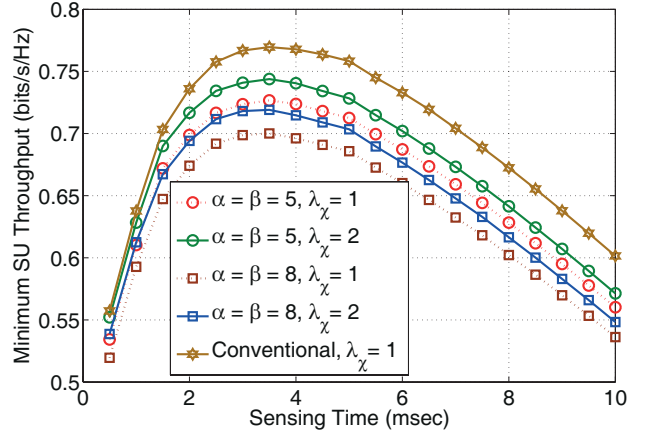} \vspace*{-6mm}
 \caption{Sensing-throughput trade-off with different channel mean power (1/$\lambda_{\chi}$) on the interference link $PT \rightarrow SR$ and different primary traffic intensities, $P_p$ = 1, $\lambda_h$ = 1.}
      \label{fig:log_norm_snr}\vspace*{-3mm}
          \end{figure}
 
\subsection{Effect of PU traffic and fading on detection performance}   
\begin{figure}
 \centering 
\includegraphics[scale=0.4]{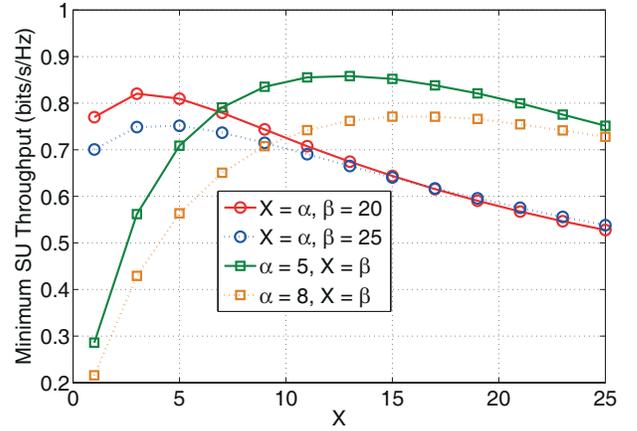} \vspace*{-4mm}
\caption{SU throughput vs. varying primary traffic, $\lambda_{\chi} = 1$, $\lambda_{h} = 1$, $P_p$ = 1, $ P_B =\frac{\alpha}{\alpha + \beta}$, $P_I = 1 - P_B$. Here, $X$ denotes the varying parameter.}
 \label{fig:traff_inten}\vspace*{-4mm}
          \end{figure}

Fig. \ref{fig:pd_pf} shows that as the primary traffic increases, the detection performance of SU degrades. Also, the condition of the channel over which the detection is performed worsens with decrease in mean power ($1/\lambda_h$) and the detection performance of SU decreases accordingly, which in turn, affects SU throughput which is discussed in Figs. \ref{fig:log_norm_snr} and \ref{fig:traff_inten}. In addition, with increase in the primary power $P_{p}$, the detection of PU becomes more reliable. Conventional model with no primary traffic performs better than the one with the primary traffic.

\subsection{Effect of PU traffic and interference link on sensing-throughput trade-off }
Fig. \ref{fig:log_norm_snr} depicts the detrimental effects of increased channel mean power (1/$\lambda_{\chi}$) of the interference link $PT \rightarrow SR$ as well as primary traffic on the average throughput of SU. As expected, SU throughput reduces with increase in the channel mean power of the interference link as SU receiver is exposed to more interference from PU transmitter. Also, increase in values of $\alpha$ and $\beta$ corresponds to increase in the primary arrival and departure rates (equivalent to decrease in mean holding time), giving rise to the increased traffic. The increase in the primary traffic leads to the drop in SU throughput.

            \begin{figure}
 \centering
 \includegraphics[scale=0.4]{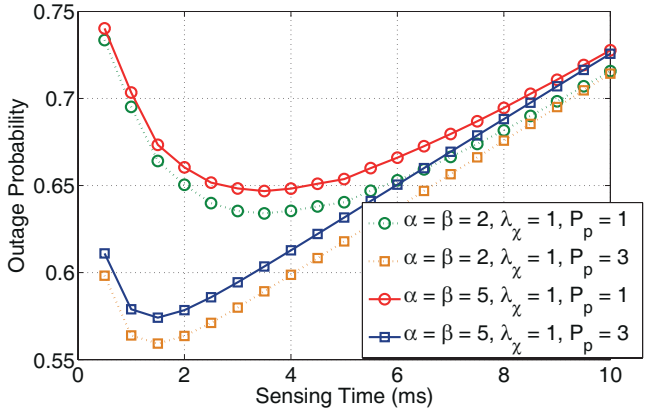}
 \caption{Variation of the outage probability vs. primary traffic, the channel power gain (1/$\lambda_{\chi}$) of the interference link and the primary transmit power, $\lambda_h$ = 1.}
  \label{fig:outage}
          \end{figure}
            \begin{figure}
\centering
   \includegraphics[scale=0.4]{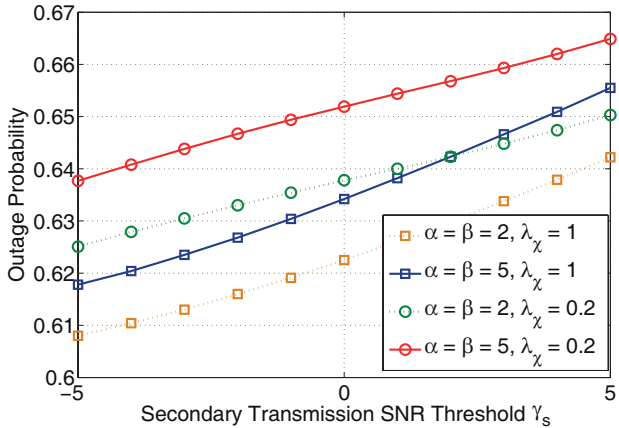} 
 \caption{Outage probability vs. secondary transmission SNR threshold $\gamma_s$, $\lambda_h$ = 1, $P_p$ = 1.}
  \label{fig:su} \vspace*{-2mm}
          \end{figure}

\subsection{Effect of varying PU traffic on SU throughput}
Fig. \ref{fig:traff_inten} shows that when either of the arrival rate $\alpha$ or the departure rate $\beta$ is kept constant and the other one is varied, there exists a non-monotonous behavior of SU throughput with varying rate. This can be relegated to the opposite effect of the probability of busy state of PU $\left(P_B = \frac{\alpha}{\alpha + \beta}\right)$ and the probability of idle state $\left(P_I = \frac{\beta}{\alpha + \beta}\right)$ on SU throughput.  As one of the rate parameters changes keeping the other one fixed, $P_B$ increases or decreases ($P_I$ decreases or increases accordingly) depending on whether the varying parameter is increasing or decreasing. For example, keeping $\beta$ fixed, increase in $\alpha$ leads to increase in $P_B$ (decrease in $P_I$). The effect of the opposite behavior of $P_B$ and $P_I$ on SU throughput can also be seen from \eqref{eq:thr_exp11}-\eqref{eq:thr_exp02} and \eqref{eq:thr}.

\subsection{Effect of PU traffic and primary interference on outage probability at SU}
From Fig. \ref{fig:outage}, it can be seen that as the sensing period increases initially, the outage probability of SU reduces due to the improved detection of PU (reducing the primary interference resulting from imperfect sensing). However, at higher sensing duration, the time allotted for transmission decreases reducing the throughput, in turn, increasing the probability of outage (decreasing the probability of successful transmission as the desired SNR constant is not satisfied). Increase in the primary traffic also increases the chances of outage. The effect of primary power $P_p$ on SU throughput is rather interesting. Though higher $P_p$ leads to more primary interference to SU, the outage probability of SU reduces with increase in $P_p$. This is due to the fact that increase in $P_p$ leads to a better detection performance (lower $\overline{P_F}$ for a given $\overline{P_D}$) as shown in Fig. \ref{fig:pd_pf}. Better detection performance also reduces the optimum sensing time, i.e., increases in the transmission period for SU, in turn, increasing the throughput of SU. This behavior can be verified from \eqref{eq:thr_exp01} and \eqref{eq:thr_exp02}. The gain obtained in the throughput because of the improved detection performance overcomes the loss due to the increased primary interference with increase in $P_p$, reducing the outage probability. 

\subsection{Effect of SNR threshold on outage probability at SU}
Fig. \ref{fig:su} shows the outage probability versus the desired transmission SNR threshold $\gamma_s$. As discussed before, the detrimental effects of the primary traffic and the increased mean interference channel power on SU throughput can be seen. As the secondary transmission SNR threshold $\gamma_s$ increases, the SNR constraint becomes tighter reducing the probability of successful transmission. This, in turn, increases the outage probability at SU. 

\bibliography{biblical}{}
\bibliographystyle{ieeetr}
\end{document}